\documentclass[11pt]{article}

\usepackage[preprint]{acl}

\usepackage{times}
\usepackage{latexsym}

\usepackage[T1]{fontenc}
\usepackage[utf8]{inputenc}

\usepackage{microtype}

\usepackage{inconsolata}

\usepackage{graphicx}

\usepackage{algorithm}
\usepackage{algorithmic}
\usepackage{amsmath}
\usepackage{amssymb}
\usepackage{multirow}
\usepackage{booktabs}
\usepackage{array}
\usepackage{enumitem}

\title{A Behavioral Fingerprint for Large Language Models: Provenance Tracking via Refusal Vectors}

\author{
  Zhenyu Xu, Victor S. Sheng \\
  Department of Computer Science, Texas Tech University \\
  \texttt{\{zhenxu, victor.sheng\}@ttu.edu}
}

\begin{document}
\maketitle
\begin{abstract}
Protecting the intellectual property of large language models (LLMs) is a critical challenge due to the proliferation of unauthorized derivative models. We introduce a novel fingerprinting framework that leverages the behavioral patterns induced by safety alignment, applying the concept of refusal vectors for LLM provenance tracking. These vectors, extracted from directional patterns in a model's internal representations when processing harmful versus harmless prompts, serve as robust behavioral fingerprints. Our contribution lies in developing a fingerprinting system around this concept and conducting extensive validation of its effectiveness for IP protection. We demonstrate that these behavioral fingerprints are highly robust against common modifications, including finetunes, merges, and quantization. Our experiments show that the fingerprint is unique to each model family, with low cosine similarity between independently trained models. In a large-scale identification task across 76 offspring models, our method achieves 100\% accuracy in identifying the correct base model family. Furthermore, we analyze the fingerprint's behavior under alignment-breaking attacks, finding that while performance degrades significantly, detectable traces remain. Finally, we propose a theoretical framework to transform this private fingerprint into a publicly verifiable, privacy-preserving artifact using locality-sensitive hashing and zero-knowledge proofs.
\end{abstract}

\section{Introduction}

The rapid proliferation of large language models (LLMs) has introduced a critical governance challenge: ensuring model provenance and protecting intellectual property in a landscape of constant modification. While the open-source release of powerful base models accelerates innovation, it also fuels an explosion of derivative works created through fine-tuning \cite{lialin2023scaling, wei2021finetuned}, quantization \cite{dettmers2022llm}, merging \cite{matena2022merging}, and other adaptations. This complex, branching lineage makes it nearly impossible to trace a model's origins, creating significant hurdles for accountability and responsible AI. For example, when a fine-tuned model generates harmful or biased content, attributing responsibility is difficult without a clear understanding of its parentage. Traditional identification methods, such as searching for architectural markers or simple watermarks, are often too brittle; they are easily erased or distorted by the very modifications they are meant to survive. This fragility underscores an urgent demand for a new class of fingerprinting technology: one that is intrinsically tied to the model's core behavior and can reliably endure extensive transformation, thus providing a stable anchor for provenance tracking.

Beyond standalone deployments, these challenges are particularly acute in the web ecosystem, where large language models are increasingly exposed as public APIs, integrated into social and content platforms, and traded on model marketplaces. When harmful outputs, copyright violations, or safety regressions occur in these web-facing systems, platforms and regulators often lack any reliable mechanism to attribute the behavior to a concrete model family or training lineage. We therefore view robust, behavior-based fingerprints as a foundational primitive for provenance-aware auditing and accountability of GenAI services on the web.

We address this challenge through a novel approach: refusal vector fingerprinting. Our key insight is that refusal mechanisms, essential components of safety alignment, create consistent directional patterns within an LLM's internal representation space. These behavioral signatures, induced by safety alignment, are deeply embedded in the model's learned parameters and remain stable across many common modifications. By extracting and analyzing these patterns, we construct distinctive fingerprints that enable accurate model attribution in both white-box settings (with direct parameter access) and, through our proposed cryptographic framework, potential black-box scenarios, providing an approach for model lineage tracking.

Our framework operates through three key stages. We first compute refusal vectors by analyzing differential activations between harmful and harmless prompts across transformer layers \cite{vaswani2017attention}, capturing the model's behavioral response patterns. These layer-specific vectors are then aggregated and normalized to produce compact, high-dimensional fingerprints that preserve discriminative information. For practical deployment, we incorporate locality-sensitive hashing and zero-knowledge proof protocols, enabling privacy-preserving verification without exposing proprietary model parameters. Model identification proceeds through similarity-based matching and hierarchical clustering, supporting both exact identification and family-level classification.

Our contributions demonstrate significant advances in model fingerprinting. We establish robustness against diverse modifications, with our refusal vector fingerprints maintaining high similarity scores across quantization, fine-tuning, adapter integration \cite{hu2021lora}, and model merging operations, though with vulnerability to alignment-breaking attacks. We prove the refusal vector fingerprint's uniqueness through comprehensive evaluation, showing low cross-family similarities among independently trained models. In large-scale identification experiments involving 76 derivative models, our method achieves high identification accuracy while significantly outperforming existing approaches. Notably, we investigate our fingerprints' behavior under alignment-breaking attacks \cite{wei2023jailbroken}, finding that while similarity decreases to approximately 0.5, this remains substantially higher than the near-zero similarities observed between unrelated model families, confirming both their foundation in safety-aligned behavior and their continued utility for forensic applications.

\section{Related Work}

The field of LLM fingerprinting has evolved from white-box intrinsic methods \cite{zeng2024huref, greshake2024reef} to black-box strategies \cite{he2024instructional, russinovich2024chain, duan2024proflingo}. While white-box methods offer high accuracy but require parameter access, black-box methods often rely on injected watermarks or active probing. Our work bridges this gap by proposing a safety-bound intrinsic fingerprint that supports both white-box extraction and privacy-preserving public verification. A comprehensive review of related work is provided in Appendix~\ref{app:related_work_extended}.

\section{Approach}

Our approach leverages the fundamental insight that refusal mechanisms in language models manifest as consistent directional patterns in their internal representation spaces. We develop a multi-stage framework that extracts these behavioral signatures, transforms them into robust fingerprints, and enables both white-box identification and black-box verification.

\subsection{Problem Formulation}

Given a language model $\mathcal{M}$ with hidden dimension $d$ and a total of $L_{\text{total}}$ transformer layers, we seek to construct a unique behavioral fingerprint $\hat{\mathbf{f}} \in \mathbb{R}^d$ that captures the model's alignment-induced refusal characteristics. Formally, our fingerprint extraction function $\Phi: \mathcal{M} \rightarrow \mathbb{R}^d$ must satisfy three key properties: \textit{uniqueness} across different model families, \textit{robustness} against common model modifications, and \textit{efficiency} in computation and comparison. Building on the theoretical foundation established by Arditi et al. \cite{arditi2024refusal}, we exploit the observation that refusal behaviors exhibit characteristic directional patterns that remain stable across model derivatives while varying distinctly between independent model families.

\subsection{Refusal Vector Computation}

Our refusal vector extraction process systematically captures the behavioral differences between harmful and harmless prompt processing across transformer layers. We compute layer-wise refusal vectors and aggregate them into a final refusal vector fingerprint. We construct two complementary prompt sets: a harmful prompt set $\mathcal{H} = \{h_1, h_2, \dots, h_n\}$ containing safety-critical queries designed to trigger refusal responses, and a harmless prompt set $\mathcal{S} = \{s_1, s_2, \dots, s_m\}$ consisting of benign instructions that should elicit helpful responses without refusal mechanisms.

For each transformer layer $\ell \in \{1, 2, \dots, L_{\text{total}}\}$, we compute the centroid representations for both prompt categories by averaging the hidden states at the final token position. The mean representations are formally defined as:
\[
\mu_\ell^{(h)} = \frac{1}{|\mathcal{H}|} \sum_{h_i \in \mathcal{H}} \mathbf{z}_\ell(h_i), \quad \mu_\ell^{(s)} = \frac{1}{|\mathcal{S}|} \sum_{s_j \in \mathcal{S}} \mathbf{z}_\ell(s_j)
\]
where $\mathbf{z}_\ell(\cdot)$ denotes the hidden representation at layer $\ell$. The refusal vector at each layer is then computed as the normalized difference between these centroids:
\[
\mathbf{r}_\ell = \frac{\mu_\ell^{(h)} - \mu_\ell^{(s)}}{\|\mu_\ell^{(h)} - \mu_\ell^{(s)}\|_2}
\]
In the rare case where the denominator is zero, indicating no discriminative signal at a given layer, that layer is excluded from the final aggregation.

To maximize discriminative power while minimizing noise, we apply a layer selection strategy. Given a selection parameter $\alpha \in (0,1]$, we define a set of excluded layers at both ends of the network. Let $k = \lfloor L_{\text{total}}(1-\alpha)/2 \rfloor$. The set of selected layers, $\mathcal{L}$, is then defined as $\{\ell \in \mathbb{Z} \mid k < \ell \leq L_{\text{total}} - k \}$. This strategy prevents overfitting to input tokenization artifacts or output projection specifics.

\subsection{Fingerprint Generation}

The final model fingerprint synthesis integrates the discriminative directional information across selected layers into a unified representation vector. We aggregate the layer-wise refusal vectors through arithmetic averaging, followed by L2 normalization to ensure unit length and facilitate reliable similarity computation:
\[
\mathbf{f} = \frac{1}{|\mathcal{L}|} \sum_{\ell \in \mathcal{L}} \mathbf{r}_\ell, \quad \hat{\mathbf{f}} = \frac{\mathbf{f}}{\|\mathbf{f}\|_2}
\]
where $\mathcal{L}$ denotes the set of selected layer indices after applying the layer selection strategy. This aggregation approach balances the contribution from different layers while reducing the impact of potential outlier directions at individual layers.

To enable comprehensive fingerprint analysis and verification, each fingerprint vector is accompanied by structured metadata $\mathcal{M} = \{d, L_{\text{total}}, |\mathcal{L}|, \alpha, |\mathcal{H}|, |\mathcal{S}|, \|\mathbf{f}\|_2, \tau\}$, encompassing the vector dimensionality, total and selected layer counts, layer selection parameter, dataset sizes, pre-normalization fingerprint magnitude, and extraction timestamp. This metadata ensures reproducibility and enables systematic comparison across different fingerprint extraction configurations.

\subsection{Model Identification and Analysis}

Our identification framework operates through similarity-based matching and clustering algorithms designed to handle both exact model identification and family-level classification. The core similarity metric employs cosine similarity to measure the angular proximity between unit vectors:
\[
 \text{sim}(\hat{\mathbf{f}}_q, \hat{\mathbf{f}}_{c_i}) = \hat{\mathbf{f}}_q \cdot \hat{\mathbf{f}}_{c_i}
\]
The candidate with the highest similarity is identified as the most likely ancestor:
\[
 \text{identified\_model} = \arg\max_{c_i} \text{sim}(\hat{\mathbf{f}}_q, \hat{\mathbf{f}}_{c_i})
\]
For large-scale analysis, hierarchical clustering (e.g., using Ward's method) can be applied to the pairwise similarity matrix of all fingerprints, grouping models into family trees based on their behavioral proximity.

\subsection{Privacy-Preserving Public Verification}

To enable public verification without revealing proprietary parameters, we propose a cryptographic framework combining Locality-Sensitive Hashing (SimHash) and Zero-Knowledge Proofs (ZKP). This allows model owners to generate a publicly verifiable proof that a fingerprint hash was correctly derived from their private model weights. Detailed protocols and mathematical formulations for this framework are presented in Appendix~\ref{app:privacy_verification}.

\section{Evaluation and Experimental Setup}

We evaluate our method on a diverse suite of 7 base models (including Llama-3.1, Qwen2.5, Phi-3, etc.) and 76 derivative models covering quantization, adapters, fine-tuning, and merges. We use cosine similarity and SimHash similarity as primary metrics, along with Top-1 identification accuracy and average margin. Detailed descriptions of datasets, models, metrics, and implementation details are provided in Appendix~\ref{app:experimental_setup}.

\section{Experiments}
\label{sec:experiments}

We conduct five comprehensive experiments to validate our refusal vector fingerprinting approach: (1) robustness evaluation against common model modifications; (2) uniqueness assessment across independently trained models; (3) large-scale identification on 76 diverse offspring models; (4) comparative analysis with state-of-the-art methods; and (5) vulnerability analysis under alignment-breaking attacks.

\subsection{Robustness Against Derivative Modifications}

We evaluate our fingerprint's robustness across seven base models spanning different architectures and scales (2B to 70B parameters) against four common modification types: quantization, adapters (using LoRA method), finetunes (using SFT), and merges. To ensure representative and high-quality offspring models, we select top-ranked derivative models based on HuggingFace's ``most likes'' sorting criterion, focusing on models with demonstrated community validation and practical usage. All base and derivative models were sourced from the HuggingFace Hub. Offspring models were selected based on the ``most likes'' sorting criterion as of September 2024. The specific HuggingFace repository identifiers for all evaluated models are listed in Appendix~\ref{app:model_details} to ensure reproducibility. Table~\ref{tab:full_comparison_final_compact} demonstrates consistent robustness across all model families, with quantization preserving the highest similarity, followed by the other three types in descending order. The strong correlation between cosine similarity and SimHash scores validates our hashing scheme for practical deployment.

To examine intra-family clustering patterns, we compute pairwise similarities for six different offspring of the Llama-3.1-8B-Instruct family. Table~\ref{tab:cos_simhash_offspring_matrix} reveals distinct clustering behavior: these four modification types exhibit high mutual similarities, forming a tight cluster. In contrast, pruning and distillation show lower similarities with other family members. These modifications are strongly destructive, as pruning removes substantial parameter information and distillation transfers knowledge without preserving original parameters, thereby significantly weakening the refusal fingerprint. This clustering pattern confirms our fingerprint's ability to distinguish between different types of model derivatives based on their parametric relationships, while also highlighting the limitations of the approach against more aggressive modifications. Detailed configuration specifications are provided in Appendix~\ref{app:model_details}.

\begin{table*}[htbp]
\centering
\small
\caption{Similarity scores for offspring models, evaluated using two metrics derived from Refusal Vector fingerprints. The table compares the direct \textbf{Cosine Similarity (CoS)} of the vectors with the approximated \textbf{SimHash Similarity (SimHS)} to validate the moderate and monotonic distortion of our SimHash scheme. Model abbreviations are: G-2B (Gemma 2B), P3M (Phi-3-mini 3.8B), Q2.5-7B (Qwen2.5 7B), L3.1-8B (Llama3.1 8B), Mxt-8x7B (Mixtral 8x7B), F-40B (Falcon 40B), L3.1-70B (Llama3.1 70B).}
\label{tab:full_comparison_final_compact}
\setlength{\tabcolsep}{4pt}
\begin{tabular}{lcccccccccccccc}
\toprule
& \multicolumn{2}{c}{\textbf{G-2B}} & \multicolumn{2}{c}{\textbf{P3M}} & \multicolumn{2}{c}{\textbf{Q2.5-7B}} & \multicolumn{2}{c}{\textbf{L3.1-8B}} & \multicolumn{2}{c}{\textbf{Mxt-8x7B}} & \multicolumn{2}{c}{\textbf{F-40B}} & \multicolumn{2}{c}{\textbf{L3.1-70B}} \\
\cmidrule(lr){2-3} \cmidrule(lr){4-5} \cmidrule(lr){6-7} \cmidrule(lr){8-9} \cmidrule(lr){10-11} \cmidrule(lr){12-13} \cmidrule(lr){14-15}
\textbf{Modification} & \textbf{CoS} & \textbf{SimHS} & \textbf{CoS} & \textbf{SimHS} & \textbf{CoS} & \textbf{SimHS} & \textbf{CoS} & \textbf{SimHS} & \textbf{CoS} & \textbf{SimHS} & \textbf{CoS} & \textbf{SimHS} & \textbf{CoS} & \textbf{SimHS} \\
\midrule
\textbf{Quantization} & 0.989 & 0.847 & 0.992 & 0.834 & 0.981 & 0.829 & 0.982 & 0.875 & 0.993 & 0.841 & 0.966 & 0.816 & 0.991 & 0.863 \\
\textbf{Adapters (LoRA)}& 0.941 & 0.793 & 0.928 & 0.769 & 0.917 & 0.751 & 0.961 & 0.766 & 0.795 & 0.642 & 0.906 & 0.778 & 0.881 & 0.724 \\
\textbf{Finetunes (SFT)}& 0.907 & 0.764 & 0.882 & 0.718 & 0.903 & 0.742 & 0.730 & 0.547 & 0.934 & 0.801 & 0.879 & 0.695 & 0.681 & 0.495 \\
\textbf{Merge}       & 0.718 & 0.591 & 0.703 & 0.556 & 0.733 & 0.603 & 0.779 & 0.500 & 0.680 & 0.534 & 0.669 & 0.521 & 0.785 & 0.612 \\
\bottomrule
\end{tabular}
\end{table*}

\begin{table*}[htbp]
\centering
\small
\caption{Cosine Similarity (CoS) / SimHash Similarity (SimHS) Matrix of \textbf{Llama-3.1-8B Offspring Models}. The SimHash values generally track the Cosine Similarity trends, though with some variance due to the dimensionality reduction inherent in the hashing process. As this is a symmetric matrix, only the lower triangular portion is shown for clarity.}
\label{tab:cos_simhash_offspring_matrix}
\renewcommand{\arraystretch}{1.2}
\begin{tabular}{lcccccc}
\toprule
 & \textbf{Unsloth-4bit} & \textbf{Nemoguard} & \textbf{UltraMedical} & \textbf{MistLlama} & \textbf{SparseLlama} & \textbf{DeepSeek-D} \\
\midrule
\textbf{Quantization (\texttt{unsloth-bnb-4bit})} & 1.000/1.000 & & & & & \\
\textbf{Adapters (\texttt{Nemoguard-8B})} & 0.950/0.781 & 1.000/1.000 & & & & \\
\textbf{Finetunes (\texttt{UltraMedical-8B})} & 0.728/0.547 & 0.688/0.578 & 1.000/1.000 & & & \\
\textbf{Merge (\texttt{MistLlama-0.1-8B})} & 0.770/0.484 & 0.726/0.484 & 0.775/0.531 & 1.000/1.000 & & \\
\textbf{Pruning (\texttt{SparseLlama-8B})} & 0.295/0.156 & 0.290/0.125 & 0.332/0.078 & 0.339/0.141 & 1.000/1.000 & \\
\textbf{Distillation (\texttt{DeepSeek-Distill-8B})} & 0.568/0.203 & 0.588/0.328 & 0.631/0.406 & 0.528/0.344 & 0.333/0.141 & 1.000/1.000 \\
\bottomrule
\end{tabular}
\end{table*}

\subsection{Uniqueness Across Independently Trained Models}

We evaluate fingerprint uniqueness by testing eight independently trained LLMs with similar architectures and scales but different training lineages. Detailed results are provided in Appendix~\ref{app:extended}, Table~\ref{tab:cos_simhash_uniqueness_matrix}. The results show that all off-diagonal pairwise similarities are consistently close to zero (average absolute similarity $<0.01$), demonstrating that our method can reliably distinguish between independently trained models.

\subsection{Large-Scale Model Family Identification}

We evaluate identification accuracy on a comprehensive test set of 76 derivative models derived from 6 distinct base families, covering four common modification types (detailed distribution in Appendix~\ref{app:model_details}). Our experimental workflow is as follows: given the pre-computed SimHash fingerprints of the 6 base model families, we compute the SimHash fingerprint for each offspring model, calculate similarity scores against all base families, rank them in descending order, and predict the top-ranked family as the source. Table~\ref{tab:exp3_accuracy_with_margin} demonstrates high (100\%) Top-1 identification accuracy across all modification types. The large average margin scores indicate clear separation between correct and incorrect candidates, demonstrating the strong discriminative power of our fingerprint method. Quantization achieves the highest confidence margin while merges achieve the lowest.

\begin{table}[htbp]
\centering
\small
\caption{Model Family Identification Performance. Our method achieved perfect Top-1 accuracy, with a large average confidence margin demonstrating a clear separation between the true parent and other candidates.}
\label{tab:exp3_accuracy_with_margin}
\begin{tabular}{lcc}
\toprule
\textbf{Modification Type} & \textbf{Top-1 Acc. (\%)} & \textbf{Avg. Margin} \\
\midrule
Quantization & 100.0 & 0.951 \\
Adapters & 100.0 & 0.842 \\
Finetunes & 100.0 & 0.823 \\
Merges       & 100.0 & 0.796 \\
\midrule
\textbf{Overall Average} & \textbf{100.0} & \textbf{0.853} \\
\bottomrule
\end{tabular}
\end{table}

\paragraph{Open-Set Identification via Thresholding.}
While our primary evaluation reports top-1 accuracy in a closed-set setting, practical deployment requires rejecting unrelated models (open-set recognition). We observe a significant ``similarity gap'' in our empirical data. \emph{Known families:} even under aggressive alignment-breaking attacks, derivative models maintain a similarity above $0.47$ (Table~\ref{tab:attack_resilience}), with standard derivatives exceeding $0.80$ (Table~\ref{tab:full_comparison_final_compact}). \emph{Unrelated families:} in contrast, independently trained models exhibit near-orthogonal refusal vectors, with pairwise similarities consistently below $0.1$ and often near zero (Table~\ref{tab:cos_simhash_uniqueness_matrix}). This distinct separation allows us to define a robust rejection threshold $\tau$ (e.g., $\tau = 0.2$). By enforcing that a valid identification requires $\max_c \text{sim}(\hat{\mathbf{f}}_q, \hat{\mathbf{f}}_{c}) > \tau$, our framework achieves a $100\%$ true negative rate on the unrelated models in Table~\ref{tab:cos_simhash_uniqueness_matrix}. This ensures that unknown models (e.g., proprietary models outside our registry) are correctly flagged as ``Unknown'' rather than being misattributed, effectively addressing the forced-choice limitation of standard top-1 classification.

\subsection{Comparison with State-of-the-Art Methods}

We compare our method against HuRef \cite{zeng2024huref} (white-box) and two black-box methods: Instructional Fingerprinting (IF) \cite{he2024instructional} and Chain \& Hash (C\&H) \cite{russinovich2024chain}. Using 48 test cases across 6 base families with various modifications, we evaluate Fingerprint Success Rate (FSR)---Top-1 identification accuracy for white-box methods and secret answer elicitation rate for black-box methods. As shown in Table~\ref{tab:exp4_sota_summary}, both white-box methods achieve 100\% FSR, significantly outperforming black-box approaches. Our advantage lies in being grounded in semantic safety behavior rather than abstract parameter distributions.

\begin{table}[htbp]
\centering
\caption{Overall Performance Comparison on Average Fingerprint Success Rate (FSR). Our FSR is defined as the Top-1 base model identification accuracy across our 48 diverse offspring models.}
\label{tab:exp4_sota_summary}
\renewcommand{\arraystretch}{1.2}
\begin{tabular}{lc}
\toprule
\textbf{Method} & \textbf{Avg. FSR (\%)} \\
\midrule
\textit{Black-Box Methods} & \\
Instructional Fingerprinting (IF) & 52.1 \\
Chain \& Hash (C\&H) & 58.3 \\
\midrule
\textit{White-Box Methods} & \\
HuRef & 100.0 \\
REEF & 100.0 \\
\textbf{Ours} & \textbf{100.0} \\
\bottomrule
\end{tabular}
\end{table}

\subsection{Vulnerability to Alignment-Breaking Attacks}

We evaluated our fingerprint's resilience against alignment-breaking attacks by testing two jailbroken variants. As detailed in Appendix~\ref{app:extended} (Table~\ref{tab:attack_resilience}), the cosine similarity drops to approximately 0.5 but remains significantly higher than the near-zero values observed between unrelated model families ($<0.1$). This confirms that while the attack weakens the refusal signal, the fingerprint retains sufficient discriminative power for provenance tracking, effectively serving as a forensic signal of safety tampering.

\subsection{Ablation Studies}

We conduct ablation studies on prompt dataset size and layer selection parameter $\alpha$. As shown in Figure~\ref{fig:ablation} in Appendix~\ref{app:extended}, our fingerprint converges rapidly with only $N=500$ prompts and remains stable across a broad range of layer selection parameters ($\alpha \in [0.3, 0.7]$), demonstrating both data efficiency and hyperparameter robustness.

\section{Discussion}
\label{sec:discussion}

Our work establishes refusal vectors as a highly effective foundation for white-box model fingerprinting. By extracting a unique feature vector derived from a model's parameters, our method demonstrates robustness in identifying derivative models through cosine similarity, even after significant modifications like fine-tuning and pruning, though with noted vulnerability to alignment-breaking attacks. The empirical results, including high identification accuracy across numerous test cases, validate our approach as a powerful tool for model provenance in white-box settings. However, the primary limitation for real-world application remains the requirement of direct model access, as suspect models are often only available as black-box APIs. This challenge motivates the development of a bridge from white-box analysis to black-box verification, for which we outline several promising directions below.

\paragraph{Comparison with intrinsic white-box baselines.}
Recent intrinsic white-box methods such as HuRef \cite{zeng2024huref} and REEF \cite{greshake2024reef} also achieve near-perfect family identification accuracy. Our approach is therefore not stronger in closed-set accuracy, but rather complementary in three important ways. First, our fingerprint is \emph{semantically bound} to safety alignment. Whereas HuRef and REEF are designed to be invariant to a wide range of benign modifications by measuring weight- or representation-level similarity, our vector is explicitly constructed from the refusal subspace. Empirically, this yields a dual-use signal: similarity remains high ($>0.9$) under common derivatives such as fine-tuning and quantization, but drops to around $0.5$ under alignment-breaking jailbreaks, while still staying far above the near-zero similarities observed between unrelated families. This behaviour turns the fingerprint into a forensic indicator of safety tampering in addition to a provenance signal, a use case that prior intrinsic methods do not explicitly target. Second, our representation is extremely lightweight: instead of maintaining activation statistics or visual artifacts, we condense each model into a single $d$-dimensional vector and, optionally, a 512-bit SimHash, making storage $O(d)$ (or $O(1)$ in the hash space) and comparison as cheap as a dot product or Hamming distance. Finally, this simple vector-and-hash pipeline is particularly amenable to cryptographic verification. While HuRef already explores zero-knowledge proofs, its circuits must encompass the full visual fingerprint generation stack, whereas our computation---from weights to refusal vector to SimHash---is closer to a standard random-hyperplane LSH pipeline, allowing more compact arithmetic circuits and making it a natural candidate for standardized, publicly verifiable model commitments.

\subsection{Pathways to Black-Box Verification}

The core challenge is to enable public verification of a computation performed on private data (the model's parameters). We propose three frameworks to achieve this, each with distinct trade-offs.

\paragraph{Cryptographically-Secured Public Artifacts.}
Inspired by the \texttt{HuRef} framework \cite{zeng2024huref}, a model owner can privately compute the fingerprint vector and transform it into a public, human-readable artifact. A zero-knowledge proof (ZKP) is then generated and published, cryptographically guaranteeing that the public artifact was honestly derived from the claimed model's parameters without revealing them. This approach offers high security and trustlessness but entails significant implementation complexity.

\paragraph{Perceptual Hashing for Similarity Estimation.}
A more direct alternative is perceptual hashing, where similar inputs produce similar outputs (i.e., hashes with low Hamming distance). The model owner would publish a perceptual hash of their private fingerprint vector. A verifier could then estimate the similarity between two models by comparing the Hamming distance of their public hashes. While computationally efficient, this method's primary challenge is security, as perceptual hashes are not inherently resistant to adversarial collision attacks.

\paragraph{Trusted Third-Party Escrow.}
A non-cryptographic, governance-based solution involves a trusted third party (TTP) acting as a secure escrow agent. Model owners would deposit their private fingerprints with the TTP, which would then perform similarity calculations in a secure environment during a dispute and issue a binding verdict. This approach is conceptually simple but relies on a centralized entity, introducing a single point of failure and potential operational overhead.

\subsection{Implications for Web-Scale Governance}

Our fingerprinting framework can be viewed as part of an emerging infrastructure layer for responsible GenAI deployment on the web. Concretely, web platforms and model marketplaces can maintain a registry of refusal-vector fingerprints for approved base models and their licensed derivatives. When a suspect web-facing service (e.g., a third-party API endpoint or a content platform using an opaque LLM backend) becomes available under unclear provenance, regulators or platform operators can require temporary white-box access to compute its refusal fingerprint and match it against the registry. High similarity scores provide evidence that the service is derived from a known, safety-aligned family, while large deviations---especially towards known jailbroken variants---serve as a forensic signal of alignment tampering. In this way, our method underpins provenance-aware auditing and accountability mechanisms for GenAI systems that shape online information flows.

\section{Limitations and Future Work}

While our approach demonstrates strong performance, we acknowledge several limitations that suggest avenues for future research.

\paragraph{White-Box Access Requirement.}
The most significant constraint of our method is its reliance on full white-box access to model parameters. This assumption is a major barrier to real-world deployment, as most proprietary models are only accessible via black-box APIs. Consequently, our approach is primarily suited for regulatory auditing (where model submission is mandated) or trusted platform verification (where platforms hold the weights), rather than for identifying unauthorized derivatives in the wild through public APIs.

\paragraph{Vulnerability to Targeted Adversarial Attacks.}
The fingerprint's robustness, though high against common modifications, has not been tested against sophisticated, targeted adversarial attacks. An adversary with knowledge of our method could potentially design attacks to specifically erase or alter the refusal vector. For instance, one could perform \textit{adversarial fine-tuning} with a curated dataset designed to rotate the refusal subspace, or employ \textit{gradient-based optimization} to directly minimize the fingerprint's magnitude in the refusal vector while preserving model utility. Investigating the feasibility and defense against such attacks is a critical direction for future work.

\paragraph{Dependence on Safety Alignment.}
While resilient to current alignment-breaking methods, the fingerprint's reliance on refusal behavior could be a vulnerability if more advanced jailbreaking techniques emerge. Future attacks might not just suppress refusal responses but fundamentally alter or erase the underlying refusal mechanisms without destroying general model utility, which would neutralize our fingerprint.

\paragraph{Path to Practical Black-Box Verification.}
The black-box verification frameworks we propose (SimHash, ZKP, TTP) represent theoretical constructs that outline pathways for practical deployment. Implementation involves engineering challenges, particularly for ZKPs applied to large-scale model parameters. Overcoming these scalability challenges represents an important research direction. Therefore, while our work establishes a strong foundation for white-box model fingerprinting, extending to practical black-box verification remains active ongoing research.

Future work will proceed along four main directions. First, we will focus on enhancing robustness by explicitly training against the targeted adversarial attacks discussed above. Second, we plan to extend our framework to multimodal architectures, investigating how refusal signals manifest in models processing combined text, image, and audio data. Third, integrating our method with formal verification frameworks could provide mathematical guarantees on fingerprint properties under specific assumptions. Finally, developing automated and scalable systems for fingerprint database management and verification, potentially starting with more lightweight solutions like perceptual hashing, is crucial for deploying our solution at scale and enabling efficient tracking across the rapidly expanding AI ecosystem.

\section{Conclusion}

In this work, we introduced a novel fingerprinting methodology based on refusal vectors, establishing a new paradigm for tracking language model provenance. By leveraging the behavioral patterns induced by model safety mechanisms, our approach generates highly distinctive and robust fingerprints that remain stable across common modifications, including fine-tuning, quantization, and merging. Our extensive experimental validation confirms the practical effectiveness of this method, achieving high identification accuracy in our extensive test suite on a large and diverse set of derivative models while demonstrating low similarity between unrelated model families. Crucially, our analysis of the fingerprint's behavior under alignment-breaking attacks reveals a dual-use capability: the drop in similarity serves as a strong forensic signal for detecting safety tampering, while the residual similarity remains high enough to distinguish the model from unrelated families, thus preserving its core fingerprinting function. The proposed cryptographic framework lays a critical foundation for transforming this white-box technique into a practical, privacy-preserving black-box verification system. Ultimately, this work represents a significant step toward building a more transparent and accountable AI ecosystem, where intellectual property is protected and the lineage of powerful models can be reliably audited. We believe this behavior-based approach will inspire further research into creating intrinsic and trustworthy model identities. In particular, when these models are deployed at web scale as APIs, platform components, or marketplace assets, refusal-vector fingerprints offer a concrete infrastructural primitive for provenance-aware and accountable use of GenAI.

\bibliography{custom}

\appendix
\section{Extended Related Work}
\label{app:related_work_extended}

The field of large language model (LLM) fingerprinting has seen rapid development, driven by the need to protect intellectual property and ensure model provenance. Methodologies can be broadly categorized by the level of access required (white-box vs. black-box) and the approach taken (intrinsic vs. injected).

\subsection{White-Box Intrinsic Fingerprinting}

White-box methods leverage internal model states for identification but require full access to the model's parameters or representations. HuRef proposes a method to create human-readable visual fingerprints from the stable vector directions of a model's parameters, using Zero-Knowledge Proofs to verify ownership without revealing the weights themselves \cite{zeng2024huref}. Another approach, REEF, focuses on comparing the similarity of intermediate layer outputs between a suspect and a victim model using Centered Kernel Alignment (CKA) \cite{kornblith2019similarity} to determine their lineage \cite{greshake2024reef}. While effective, both methods are fundamentally limited to scenarios where white-box access is feasible.

\subsection{Injected Watermarks for Black-Box Verification}

To address the limitations of white-box access, injected methods embed a verifiable signal into the model that can be queried via a standard API. Instructional Fingerprinting (IF) pioneers this by using instruction tuning \cite{wei2021finetuned} to implant a secret trigger-response pair as a backdoor \cite{he2024instructional}. To counter forgery, where an attacker could implant their own fingerprint, Chain \& Hash introduced a cryptographically-linked chain of question-answer pairs, making it computationally infeasible to generate conflicting fingerprints and thus providing unforgeable ownership proof \cite{russinovich2024chain}. Addressing the vulnerability of explicit backdoors, Implicit Fingerprints (ImF) uses steganography and Chain-of-Thought reasoning to create semantically coherent trigger-response pairs that are indistinguishable from the model's normal behavior, enhancing stealth \cite{li2025imf}.

\subsection{Black-Box Behavioral Analysis}

Active fingerprinting techniques probe a black-box model with specifically crafted queries to elicit identifying responses. The ProFLingo method generates queries inspired by adversarial examples, designed to provoke a target response from a derivative model that would be highly unlikely from an unrelated one \cite{duan2024proflingo}. Similarly, LLMmap functions like a network scanner, sending a small number of carefully designed queries to identify a specific model version from a large set, positioning model fingerprinting as a reconnaissance tool in security assessments \cite{narayanan2024llmmap}. In contrast, passive methods analyze the statistical properties of a model's output. Some work treats the problem as one of author attribution, using linguistic features and n-grams to identify the unique stylistic ``fingerprint'' of an LLM family \cite{povincze2025stylistic}.

\subsection{Attacks and Defenses in Fingerprinting}

As fingerprinting techniques evolve, so do the attacks designed to remove them. The model merging technique, which combines multiple models, poses a unique threat by diluting fingerprints. MergePrint was specifically designed to be robust against this, embedding the fingerprint in a way that survives the merging process \cite{song2024mergeprint}. Direct removal attacks have also been developed. MEraser is an explicit ``anti-virus'' tool that uses a two-stage fine-tuning process to first break the trigger-response association of a backdoor fingerprint and then restore the model's general capabilities, effectively erasing the watermark \cite{he2025meraser}. To counter collusion attacks, where users could share and filter out known fingerprints, Perinucleus Sampling was developed to embed tens of thousands of unique fingerprints by sampling responses from the low-probability ``perinucleus'' region of the output distribution, making large-scale removal impractical \cite{lee2024perinucleus}.

\section{Privacy-Preserving Verification Details}
\label{app:privacy_verification}

\paragraph{Locality-Sensitive Hashing.}
Following the approach established by HuRef \cite{zeng2024huref}, we employ Random Projection-based SimHash \cite{charikar2002similarity}, a specific type of locality-sensitive hashing (LSH) technique that preserves vector similarity in the hash domain. Given a $d$-dimensional refusal vector $\mathbf{v}$, we generate a $k$-bit hash through random hyperplane projections. Specifically, we pre-generate a matrix $\mathbf{R} \in \mathbb{R}^{k \times d}$ of random hyperplanes, where the $i$-th hash bit $h_i$ is determined by $h_i = \mathbf{1}[\mathbf{r}_i \cdot \mathbf{v} \geq 0]$. The resulting hash $H(\mathbf{v})$ is the integer representation of this k-bit string.

For similarity computation between hashes $H_1$ and $H_2$, we calculate the Hamming distance $d_H(H_1, H_2)$ and convert it to a raw similarity score $S_{\text{raw}} = 1 - d_H/k$. To align with cosine similarity ranges, we apply the transformation $S_{\text{adj}} = 2(S_{\text{raw}} - 0.5)$, which re-centers the baseline from 0.5 to 0 for uncorrelated vectors.

\paragraph{Zero-Knowledge Proof Framework.}
To ensure cryptographic integrity without revealing private parameters, we implement a three-phase zero-knowledge proof protocol \cite{goldwasser1985knowledge}. The \textit{Setup Phase} compiles the entire computation pipeline—from model parameters $\mathcal{W}$ to hash $H$—into a provable arithmetic circuit, generating a private Proving Key ($\text{PK}$) and public Verification Key ($\text{VK}$). During the \textit{Proving Phase}, the model owner privately computes $\pi = \text{Prove}(\text{PK}, \mathcal{W}, H)$, generating a succinct proof that the hash was correctly derived from their model parameters. Finally, the \textit{Verification Phase} enables any party to publicly verify the proof's validity through $\text{Verify}(\text{VK}, H, \pi)$, which returns a cryptographic guarantee of authenticity without revealing the underlying model parameters. This framework enables practical deployment in scenarios where model owners must prove fingerprint ownership without exposing proprietary parameters, bridging the gap between white-box extraction and black-box verification requirements. In web settings, such public, privacy-preserving fingerprints would allow platforms to credibly commit to the identity of the models serving user requests, while enabling external auditors, civil society organizations, or regulators to verify these claims without direct access to proprietary parameters.

\section{Experimental Setup Details}
\label{app:experimental_setup}

\subsection{Datasets and Models}

We now describe the dataset construction for refusal vector computation, followed by our diverse evaluation model ecosystem spanning different architectures, modification types, and training lineages.

\subsubsection{Refusal Vector Computation Dataset.}
Following the methodology established by Arditi et al. \cite{arditi2024refusal}, we construct a balanced dataset of approximately 10{,}000 prompts. The dataset comprises 5{,}000 harmful prompts sourced from standard safety benchmarks like AdvBench \cite{zou2023universal} and JailbreakBench \cite{chao2023jailbreaking}, and 5{,}000 harmless prompts from general instruction-tuning datasets like Alpaca \cite{taori2023stanford} and ShareGPT \cite{zheng2023judging}.

\subsubsection{Evaluation Models.}
We evaluate our fingerprinting approach across a diverse ecosystem of language models spanning different architectures, scales, and training lineages. Our base model collection includes Meta Llama-3.1 (8B, 70B), Qwen2.5 (7B), Gemma-2B, Microsoft Phi-3-mini (3.8B), Falcon-40B, and Mixtral-8x7B, representing the current state-of-the-art in open-source language modeling. These models employ various architectural innovations including grouped query attention \cite{ainslie2023gqa}, rotary position embeddings \cite{su2021roformer}, and mixture-of-experts designs \cite{fedus2021switch}. These models are primarily sourced from HuggingFace, with derivative models selected based on community validation using the ``most likes'' ranking criterion to ensure quality and practical relevance. This selection strategy ensures the use of high-quality, relevant models for evaluation. To assess robustness against common model modifications, we systematically generate derivative models through four types of modifications: quantization, adapters (LoRA), fine-tuning (SFT), and model merging. Additionally, we evaluate fingerprint uniqueness using independently trained models from different organizations with similar architectures but distinct training pipelines, and test resilience against alignment manipulation using jailbroken models with modified refusal behaviors.

\subsection{Evaluation Metrics}

Our evaluation framework employs a comprehensive set of metrics to assess both the accuracy and robustness of our fingerprinting approach across diverse experimental scenarios.

\paragraph{Similarity Metrics.}
We utilize two complementary similarity measures to quantify fingerprint correspondence. \textit{Cosine Similarity} serves as our primary metric, measuring the angular similarity between refusal direction vectors in the high-dimensional representation space. \textit{SimHash Similarity} provides a dimensionality-reduced approximation that enables efficient large-scale comparisons while preserving the essential discriminative properties of the original vectors.

\paragraph{Identification Performance.}
\textit{Top-1 Identification Accuracy} quantifies the percentage of cases where our method correctly identifies the true source model family from a candidate set, based on the highest similarity score. This metric directly measures the practical effectiveness of our fingerprinting approach for model provenance tracking.

\paragraph{Confidence Assessment.}
\textit{Average Margin} captures the discriminative power of our method by measuring the gap between the top-1 and top-2 similarity scores. Larger margins indicate stronger confidence in identification decisions and better separation between correct and incorrect candidates.

\paragraph{Comparative Evaluation.}
For direct comparison with existing fingerprinting methods, we adopt the \textit{Fingerprint Success Rate (FSR)} metric, following the evaluation protocol established by prior work \cite{zeng2024huref}. Our FSR corresponds to the Top-1 identification accuracy, enabling standardized performance comparisons across different fingerprinting approaches.

\subsection{Implementation Details}

Our implementation follows established best practices for neural model analysis. We target the middle 90\% of transformer layers for refusal vector computation, extracting hidden states from the final token position. All direction vectors undergo L2 normalization. For hashing, we employ 512-bit SimHash and adjust raw scores via $S_{adj} = 2(S_{raw} - 0.5)$ to align with cosine similarity. We utilize bfloat16 precision for all experiments. For identification tasks, candidates are ranked in descending order of similarity, and the average margin ($\text{sim}_{\text{top-1}} - \text{sim}_{\text{top-2}}$) is used to quantify identification confidence.

\section{Extended Results and Discussion}
\label{app:extended}

\subsection{Uniqueness Across Independently Trained Models (Table)}

Table~\ref{tab:cos_simhash_uniqueness_matrix} presents the pairwise similarity matrix for independently trained models, supporting the uniqueness claims in Section~\ref{sec:experiments}.

\begin{table*}[h!]
\centering
\small
\caption{Pairwise Cosine Similarity (CoS) / SimHash Similarity (SimHS) Matrix for \textbf{Independently Trained LLMs} (7B Scale). The consistently low off-diagonal values confirm fingerprint uniqueness across model families. As this is a symmetric matrix, only the lower triangular portion is shown for clarity.}
\label{tab:cos_simhash_uniqueness_matrix}
\begin{tabular}{lcccccccc}
\toprule
\textbf{Model} & \textbf{Mistral} & \textbf{Granite} & \textbf{InternLM2} & \textbf{Baichuan2} & \textbf{Qwen2.5} & \textbf{Llama2} & \textbf{BLOOM} & \textbf{Llama3.1} \\
\midrule
\textbf{Mistral}   & 1.000/1.000 & & & & & & & \\
\textbf{Granite}   & -0.010/-0.031 & 1.000/1.000 & & & & & & \\
\textbf{InternLM2} & 0.025/0.031 & -0.013/-0.031 & 1.000/1.000 & & & & & \\
\textbf{Baichuan2} & 0.001/0.094 & 0.004/-0.094 & 0.000/-0.063 & 1.000/1.000 & & & & \\
\textbf{Qwen2.5}    & 0.000/-0.109 & 0.016/-0.016 & 0.020/-0.047 & -0.004/-0.172 & 1.000/1.000 & & & \\
\textbf{Llama2}    & 0.008/0.094 & -0.018/0.125 & -0.003/-0.094 & 0.031/-0.125 & -0.014/-0.141 & 1.000/1.000 & & \\
\textbf{BLOOM}     & -0.002/-0.063 & 0.015/0.063 & -0.014/0.063 & 0.019/-0.063 & 0.001/0.078 & 0.017/-0.094 & 1.000/1.000 & \\
\textbf{Llama3.1}  & -0.013/-0.078 & -0.003/0.047 & 0.030/0.078 & 0.022/0.016 & -0.011/0.063 & 0.007/0.016 & 0.002/-0.078 & 1.000/1.000 \\
\bottomrule
\end{tabular}
\end{table*}

\subsection{Intra-Family Clustering (Table)}

Table~\ref{tab:cos_simhash_offspring_matrix} in the main text provides the pairwise similarities for Llama-3.1-8B offspring models, illustrating the clustering behavior described in Section~\ref{sec:experiments}.

\subsection{Vulnerability to Alignment-Breaking Attacks (Table)}

Table~\ref{tab:attack_resilience} provides the specific similarity scores for the jailbroken models discussed in Section~\ref{sec:experiments}.

\begin{table}[htbp]
\small
\centering
\caption{Fingerprint Similarity for Models Under Alignment-Breaking Attacks.}
\label{tab:attack_resilience}
\begin{tabular}{lcc}
\toprule
\textbf{Attacked Variant Model} & \textbf{CoS}& \textbf{SimHS} \\
\midrule
\texttt{failspy/abliterated-v3} & 0.4746 &0.2564 \\
\texttt{cooperleong00/Jailbroken} & 0.5020 &0.2872\\
\bottomrule
\end{tabular}
\end{table}

\subsection{Ablation Studies (Figure)}

\begin{figure*}[htbp]
  \centering
  \includegraphics[width=\textwidth]{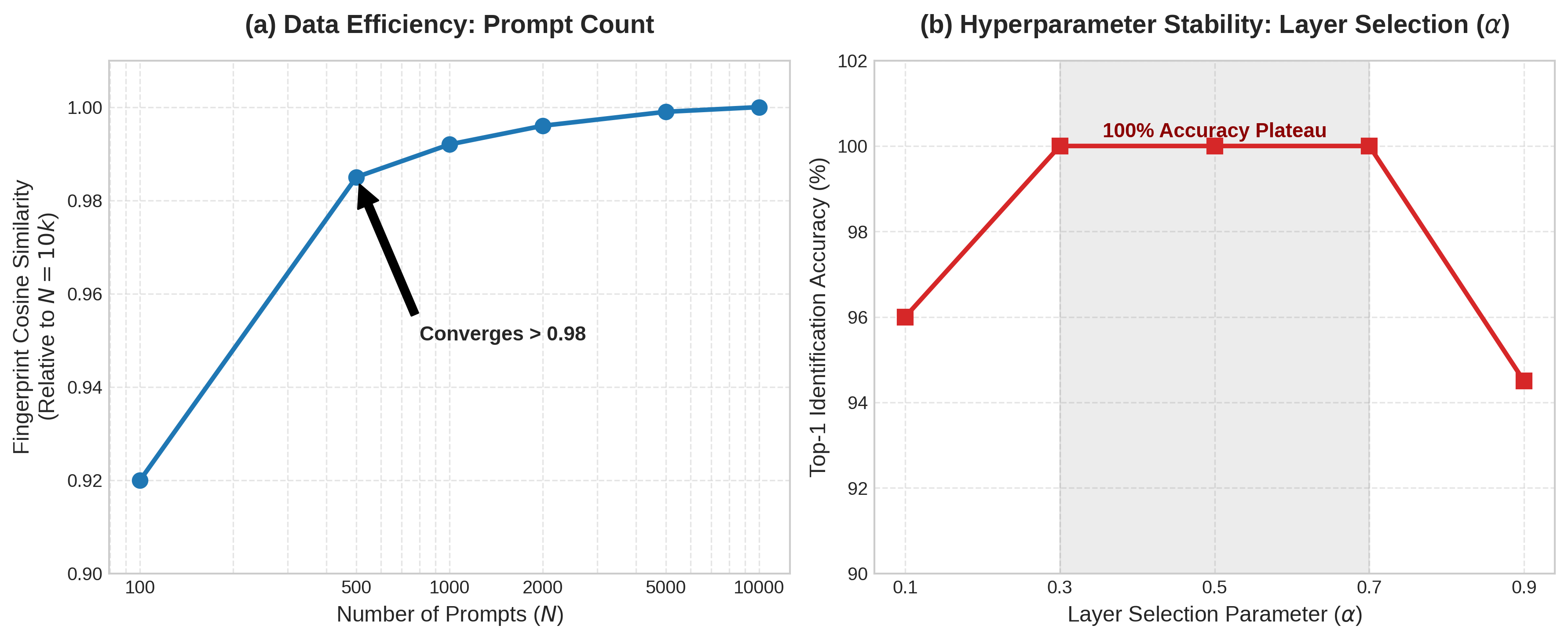}
  \caption{Ablation studies on data efficiency and hyperparameter sensitivity. (a) Fingerprint stability converges rapidly; using only $N=500$ prompts achieves $>0.98$ cosine similarity relative to the full 10k baseline, demonstrating extreme extraction efficiency. (b) Identification performance remains robust (100\% accuracy) across a wide range of layer selection parameters ($\alpha \in [0.3, 0.7]$), indicating that the method does not require fine-grained tuning.}
  \label{fig:ablation}
\end{figure*}

Figure~\ref{fig:ablation} visualizes the results of our ablation studies on data efficiency and hyperparameter sensitivity.

\section{Model Details}
\label{app:model_details}

Table~\ref{tab:hf_model_ids} lists the exact HuggingFace repository identifiers for the models used in our experiments.

\begin{table}[h]
\centering
\scriptsize
\setlength{\tabcolsep}{3pt}
\renewcommand{\arraystretch}{0.95}
\caption{HuggingFace Repository Identifiers for Base and Offspring Models.}
\label{tab:hf_model_ids}
\begin{tabular}{ll}
\toprule
\textbf{Model Shorthand} & \textbf{HuggingFace Repository ID} \\
\midrule
\multicolumn{2}{l}{\textit{Base Models}} \\
Llama-3.1-8B & \texttt{meta-llama/Meta-Llama-3.1-8B-Instruct} \\
Llama-3.1-70B & \texttt{meta-llama/Meta-Llama-3.1-70B-Instruct} \\
Qwen2.5-7B & \texttt{Qwen/Qwen2.5-7B-Instruct} \\
Gemma-2B & \texttt{google/gemma-2b-it} \\
Phi-3-mini & \texttt{microsoft/Phi-3-mini-4k-instruct} \\
Falcon-40B & \texttt{tiiuae/falcon-40b-instruct} \\
Mixtral-8x7B & \texttt{mistralai/Mixtral-8x7B-Instruct-v0.1} \\
\midrule
\multicolumn{2}{l}{\textit{Llama-3.1-8B Derivatives (Offspring)}} \\
Unsloth-4bit & \texttt{unsloth/Meta-Llama-3.1-8B-Instruct-bnb-4bit} \\
Nemoguard & \texttt{nvidia/Llama-3.1-Nemoguard-8B} \\
UltraMedical & \texttt{KaijuLLM/UltraMedical-Llama-3.1-8B} \\
MistLlama & \texttt{Weyaxi/MistLlama-v1} \\
SparseLlama & \texttt{neuralmagic/SparseLlama-3.1-8B-2of4} \\
DeepSeek-Distill & \texttt{deepseek-ai/DeepSeek-R1-Distill-Llama-8B} \\
\midrule
\multicolumn{2}{l}{\textit{Attacked Variants}} \\
Abliterated-v3 & \texttt{failspy/llama-3-8b-instruct-abliterated-v3} \\
Jailbroken & \texttt{cooperleong00/Jailbroken-Llama-3-8b} \\
\bottomrule
\end{tabular}
\end{table}

\end{document}